# Theoretical study of transport properties of $B_{40}$ and its endohedral borospherenes in single-molecule junctions

*Chengbo Zhu and Xiaolin Wang\**

*Spintronic and Electronic Materials Group, Institute for Superconducting and Electronic Materials, Australian Institute for Innovative Materials, University of Wollongong, North Wollongong, New South Wales 2500, Australia*

ABSTRACT $C_{60}$ fullerene has been studied extensively, as it is considered to be a good candidate for building single-molecule junctions. Here, we theoretically demonstrate that the conductance of single-molecule junctions based on a newly discovered molecule, borospherene ($B_{40}$), is comparable to that for the $C_{60}$-based junction with its more delocalized π electrons. The charge injection efficiency in the $B_{40}$-based junction is improved, as up to 7 atoms in direct contact with the electrode are possible in the Au-$B_{40}$-Au junction. Interestingly, a higher number of atoms in direct contact with the electrode does not result in a higher number of conduction channels because of the unique chemical bonding in the $B_{40}$ molecule, without two-center two-electron bonds. The transport properties of Au-$B_{40}$-Au junctions can be proved by doping. With a Ca, Sr, or Y atom encapsulated into the $B_{40}$ cage, the conductance at zero bias increases significantly. Moreover, our calculations show that the lowest unoccupied molecular orbital dominates the low-bias transport, as the thermopower in these junctions is negative. Our study indicates that $B_{40}$ is an attractive new platform for designing highly conductive single-molecule junctions for future molecular circuits.

I. INTRODUCTION

Allotropy, where the atoms of an element are bonded together in different manners, can lead to distinctive electronic properties from the different atomic structures. The most famous example is Carbon, which has many allotropes, from 3-dimensional diamond and graphite to 0-dimensional buckminsterfullerene.[1] While diamond is an electrical insulator, buckminsterfullerene is conductive and can become a superconductor after doping.[2-7] Although buckminsterfullerene was discovered in 1985, its properties are still being intensively studied because of its promising applications in molecular electronics.[8-24] Meanwhile, searching for fullerenes made of materials other than Carbon has been an intriguing topic for researchers.[25] The natural place to look is at the adjacent element to Carbon in the periodic table, Boron. There have been various theoretical works predicting the existence of all-boron fullerene.[26-38] Recently, the first all-boron fullerene-like cage cluster molecule, $B_{40}$, was observed experimentally.[39] It is well known that pure Boron is an electrical insulator at room temperature, so the question of whether this newly discovered all-boron fullerene molecule exhibits exotic electronic properties on the mesoscopic scale and has potential applications in the field of molecular electronics remains elusive. Furthermore, can the electronic properties of $B_{40}$ be tuned? The goal of this work is to study these two questions from a theoretical point of view.

A number of C-based materials, such as carbon nanotubes, graphene sheets, and carbon nanoribbons, are promising components for future nanoelectronics[40] due to their unique transport properties and versatile applications.41 While the injection and the collection of charges between these graphitic structures and external metallic leads are controllable,[42,43] it is more challenging to form a stable contact in single-molecule junctions. For $C_{60}$-based junctions, various contact geometries have been proposed and studied both theoretically and experimentally.[10-22] Understanding the transport characteristics of $C_{60}$ fullerene bonded between metal electrodes is of fundamental importance, because it is thought to be a good candidate to build highly conductive single-molecule junctions. Based on scanning tunnelling microscopy (STM), different contact geometries and electrode materials have been constructed and used to measure the conductance of $C_{60}$ fullerene.[9-13,15,44] Together with the theoretical calculations, the reported values for the conductance vary between ~$10^{-4}$ $G_0$ and 1 $G_0$.[16-19,22,45,46] These values are scattered over more than 3 orders of magnitude. Searching for highly conductive single-molecule junctions with stable contacts remains a challenge.

Recently, the first all-boron fullerene-like cage cluster molecule, $B_{40}$, with an extremely low electron binding energy has been observed experimentally.[39] Interest in the novel properties of the $B_{40}$ molecule and its endohedral metalloborospherenes[47,48] has been growing. This has encouraged the further exploration of $B_{40}$'s potential application in molecular electronics. There is an urgent need for investigations to demonstrate if the junctions based on $B_{40}$ have advantages over $C_{60}$-based junctions and, in turn, make $B_{40}$ a good candidate for future molecule-based electronics.

The contact geometry of the $B_{40}$-based junction is expected to be more stable than that of the $C_{60}$-based junction, owing to the atomic structure of the $B_{40}$ molecule. It is easy to form a contact between one of the hexagonal or heptagonal faces of the molecule and the electrode. Six or seven boron atoms would form a direct contact with the metallic electrode, leading to a higher injection rate of charges.

Furthermore, unlike the $C_{60}$ molecule, there is no localized two-center two-electron bond (2c-2e) on the $B_{40}$ molecule. It is well known that the degree of delocalization of the π electrons plays an important role in the electrical conduction in the $C_{60}$-based molecular junction. All of the π electrons on the $B_{40}$ molecule are 5c-2e, 6c-2e, or 7c-2e bonds.[39] With more delocalized electrons, $B_{40}$ as a highly conductive molecule is an ideal candidate. So far, the transport properties of the $B_{40}$ molecular junction have not been studied.

In this article, our aim is to study the transport properties of single-molecule junctions based on the $B_{40}$ molecule and its endohedral borospherenes. Gold electrodes have been used in our calculations. The results show that the conductance of the Au-$B_{40}$-Au junction is comparable to that of the Au-$C_{60}$-Au junction. Two contact geometries have been simulated: one is formed by using the hexagonal faces to couple with the electrodes; and the other is formed by using the heptagonal faces to couple with the electrodes. In the contact regime, the $B_{40}$ molecular junction is more conductive when the heptagonal face is used to couple the electrodes, because of the greater number of B atoms in direct contact with the electrodes. Furthermore, we have studied the thermopower of the $B_{40}$ molecule. We find that the low-bias transport is mainly dominated by the lowest unoccupied molecular orbital (LUMO) of the molecule. Also, the doping effect is significant in tuning the transport properties. Our results reveal that $B_{40}$ provides a new platform for designing highly conductive single-molecule junctions for future molecular circuits.

## II. COMPUTATIONAL METHODS

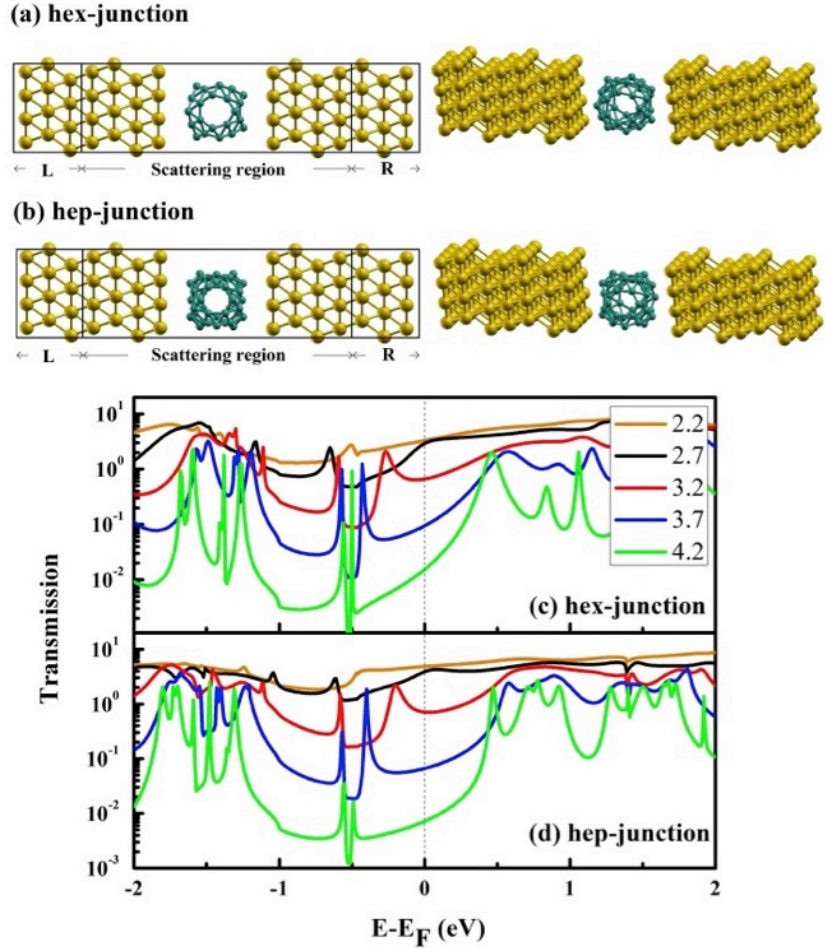

**Figure 1.** Au-$B_{40}$-Au junctions with different surface coupling to the electrodes viewed from different angles: (a) hex-junction (hexagonal faces pointed at electrodes) and (b) hep-junction (heptagonal faces pointed at electrodes). Transmission as a function of energy at zero bias for the two geometries: (c) hex-junction, and (d) hep-junction, with different B-Au distances from 2.2 Å to 4.2 Å.

The density functional theory (DFT)-based non-equilibrium Green's function (NEGF) formalism has been employed to calculate the transport properties.[49]

The systems studied can be divided into three regions: the central region, the left electrode, and the right electrode, as shown in Figure 1(a) and (b). The electronic structure for the central region was calculated using SIESTA.[50] The single $B_{40}$ molecule was relaxed first. Then, the molecular junctions were constructed by structures comprising an 8-layer slab of Au (111) in a 4×4 representation and the relaxed $B_{40}$ molecule. A 1×1×100 Monkhorst-Pack k-point mesh was used. The $B_{40}$ molecule, which was sandwiched in the junction with 4 Au layers on each side was optimized again until the forces on all the $B_{40}$ atoms were smaller than 0.03 eV/Å. The subsequent transmission calculations were carried out using TranSIESTA.[49] The zero-bias conductance G can be expressed as[51]

$$G = G_0 \mathcal{T}(E), \tag{1}$$

where $\mathcal{T}(E)$ is the transmission function, $G_0 = 2e^2/h$. Within the standard NEGF formalism, the transmission function is given by

$$\mathcal{T}(E) = \text{Tr}[\Gamma_L(E)G(E)\Gamma_R(E)G^+(E)], \tag{2}$$

where the retarded Green's function $G(E)$ is

$$G(E) = [(E + i\eta)S - H - \Sigma_L(E) - \Sigma_R(E)]^{-1}, \tag{3}$$

with $S$ and $H$ being the overlap and Hamiltonian matrices of the central region, respectively. The electrode-coupling effect is evaluated by the self-energies as

$$\Gamma_{L/R}(E) = i[\Sigma_{L/R}(E) - \Sigma_{L/R}^{+}(E)]. \qquad (4)$$

The generalized gradient (GGA) Perdew-Burke-Ernzerhof (PBE) approximation was used for exchange-correlation.[52] A single-zeta plus polarization basis set for Au atoms and double-zeta plus polarization basis set for $B_{40}$ atoms were employed. The mesh cut-off was chosen as 300 Ry. The individual transmission coefficients were calculated using Inelastica.[53,54]

To simulate the stretching of the contacts, we started with a geometry in which the molecule is positioned between two gold electrodes with flat surfaces. Due to the atomic structure of the $B_{40}$ molecule, two contact geometries can be formed: with two hexagonal or two heptagonal faces being coupled with the electrodes. In the following, we refer to the two types of junctions as hex/hep-junctions, in which the hexagonal/heptagonal face is used to couple with the electrodes, respectively. The original distance between the surface of a gold electrode and the nearest edge atoms (atoms in the hep/hex face) of the inserted $B_{40}$ molecule was set to 1.7 Å. The hex-junction and the hep-junction studied in our calculations are shown in Figure 1(a) and (b). Then, the gold electrodes were separated stepwise from the molecule (in steps of ~ 0.5 Å), and the junction geometry was relaxed at every step. This protocol was repeated until the junction was broken and the molecule lost contact with the electrodes. During the stretching of the contact, the $B_{40}$ molecule moves up and down to form a stable geometry.

In this work, the binding energy, $E_b$, is calculated by using Eq. (5),

$$E_b = E_T(B_{40} + Au) - [E_T(B_{40}) + E_T(Au)], \qquad (5)$$

where $E_T$ is the total electronic energy, (B$_{40}$+Au) represents the B$_{40}$-based junction, (B$_{40}$) represents a single B$_{40}$ molecule, and (Au) in Equation (5) represents the junction without the B$_{40}$ molecule inserted. A negative binding energy thus corresponds to a stable system.

Another transport property of interest in this work is the thermopower (*S*; also known as the Seebeck coefficient). At zero applied bias voltage, *S* can be calculated by

$$S = -\frac{\pi^2 k_B^2 T}{3e} \frac{\mathcal{T}'(E_F)}{\mathcal{T}(E_F)}, \qquad (6)$$

where $\mathcal{T}(E_F)$ is the transmission function at the Fermi level ($E_F$), and the prime denotes the derivative with respect to energy, $k_B$ is the Boltzmann constant, *T* is the temperature (*T* = 300 K in our calculations), and *e* is the charge of the electron. The sign of *S* can be used to deduce the nature of the charge carriers in molecular junctions: a positive *S* results from hole transport through the highest occupied molecular orbital (HOMO), whereas a negative *S* indicates electron transport through the LUMO.[53]

III. RESULTS AND DISCUSSION

**A. Transport properties of B$_{40}$-based junction.** To achieve as many conductive channels as possible, two electrodes with ideal surfaces were considered in our calculations, as shown in Figure 1 (a) and (b). The transmission for the two types of the junctions with various B-Au distances is depicted in Figure 1(c) and (d). For the two types of junctions, the zero-bias conductance, which is determined by the transmission at the $E_F$, increases exponentially while the B-Au distance decreases. When the B-Au distance is larger than 3.2 Å, the transmission shows peaks related to the molecular energy levels of the B$_{40}$'s orbitals. The closer the B-Au distance is, the smaller the HOMO-LUMO gap will be, as the coupling between the molecule and the electrodes becomes stronger. When the B-Au distance is smaller than 2.7 Å, however, the coupling between the electrodes and the molecule is so strong that the HOMO and LUMO peaks are broadened significantly, resulting in transmission without pronounced peaks around the $E_F$.

The zero-bias conductance of the $B_{40}$-juntion with B-Au distance of 2.2 Å is 4.86 $G_0$ and 3.31 $G_0$ for the hep-junction and hex-junction, respectively; with B-Au distance of 2.7 Å, the zero-bias conductance is 3.9 $G_0$ and 2.92 $G_0$ for the hep-junction and hex-junction, respectively. Conductances above 1 $G_0$ have been reported in theoretical studies of $C_{60}$ junctions with Al,[18,55] Au,[19,22] and Cu[21] electrodes, when the leads are similar to ideal surfaces, i.e., with high Au-$C_{60}$ coordination. For $C_{60}$ junctions with different contact geometries where the electrodes are made of Au, the relaxed C-Au distances fall between 2.15-2.45 Å.[20,22,56] It is clear that the conductance of a $B_{40}$-based junction is comparable to that of a $C_{60}$-based junction with similar molecule-electrode distances.

The conductance is mainly dependent on two factors, the charge injection rate and the ability to scatter electrons.[13] The first factor is generally dependent upon the contact geometry. Since electrodes with ideal surfaces are used in our calculations, the charge injection rate is maximized: for hep (hex)-junctions, 7 (6) B atoms would have direct contact with the metallic electrode. The bottleneck is therefore the intrinsic ability to scatter electrons.

The conductance of a molecular junction is dependent on the molecular length,[57-59] and it usually decreases exponentially as the molecular length increases. The tunneling decay constant of a series of alkane diamines, for example, is 0.91 ±0.03 per methylene group.[59] The nucleus-to-nucleus diameter of $C_{60}$ is ~7 Å. The $B_{40}$ molecule has 20 fewer atoms than the $C_{60}$ molecule, and thus, it has a smaller diameter. The distance between the two furthest atoms on opposite heptagonal faces of a single $B_{40}$ molecule, according to the results of our calculation, is ~5.57 Å; the distance between the two furthest atoms on the opposite hexagonal faces is ~5.2 Å. With a larger charge injection rate and smaller diameter, however, the conductance of a $B_{40}$-based junction is not remarkably higher than that of a $C_{60}$-based junction.

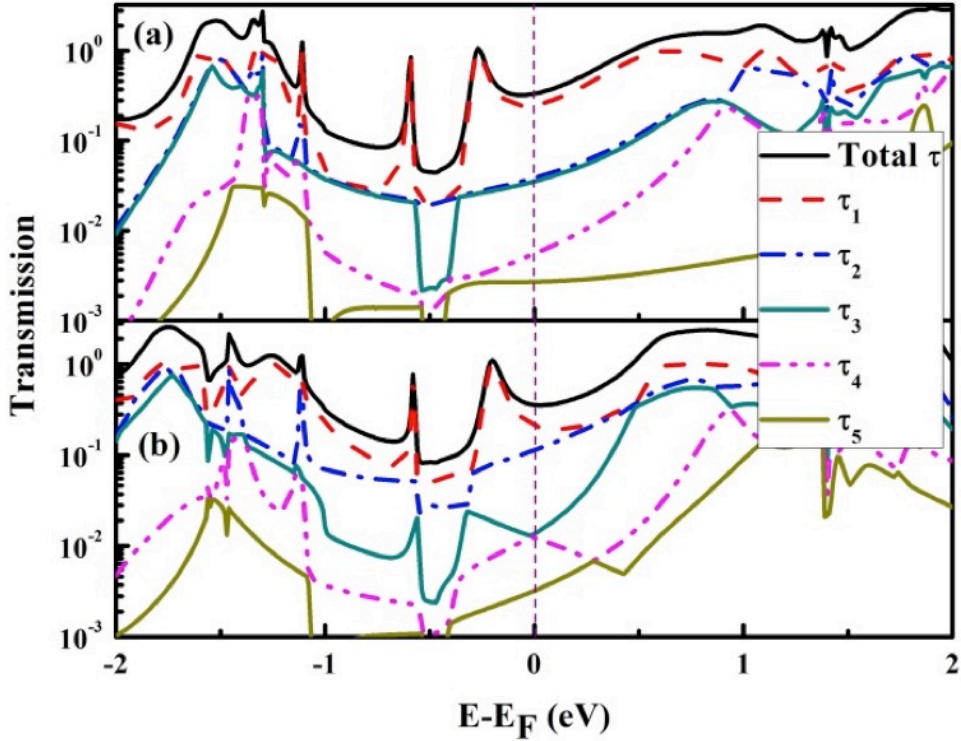

**Figure 2.** Transmission as a function of energy for a B-Au distance of 3.2 Å for the two geometries: (a) hex-junction and (b) hep-junction. The solid black line corresponds to the total transmission, while the other lines correspond to the contributions of the individual transmission coefficients as functions of energy.

To explain this, the transmission curves of Au-$B_{40}$-Au junctions for the two geometries at the B-Au distance of 3.2 Å and their channel decompositions are shown in Figure 2. (Transmission curves of Au-$B_{40}$-Au junctions with individual conduction channels at different B-Au distances are shown in Figures S1-S18 in the Supporting Information.) As we can see from Figure 2, the number of conduction channels which contribute significantly to the conductance is smaller than the number of B atoms in direct contact with the electrode. For a single-molecule junction based on a π-conjugated molecule or $C_{60}$ fullerene, the number of conduction channels is generally given by the number of C atoms bonded to the surface of the electrode, because each C atom provides one π-channel. In the $B_{40}$ junction, even when the B-Au distance is as close as 1.7 Å, the contributions to the total transmission from the fifth and sixth conduction channels are negligible (Figures S1-S18). This is due to the unique chemical bonding in the all-boron fullerene: on average, each boron atom contributes 0.6 electrons to the π bonding,[39] which is responsible for the conductance in the $B_{40}$-molecule junction. Still, with a similar molecule-electrode contact distance, the $B_{40}$-based junction is more conductive compared with the $C_{60}$-based junction. The conductance can be further improved by doping.

To characterize the junction during the stretching process, various quantities other than conductance were calculated, such as the binding energy, Mulliken charges, and Seebeck coefficient, as shown in Figure 3. As can be seen from Figure 1(c) and (d), the conductance drops exponentially during the stretching process. It is more straightforward to see the trend in Figure 3(a), which shows the transmission as a function of the $B_{40}$-Au distance for both hex- and hep-junctions.

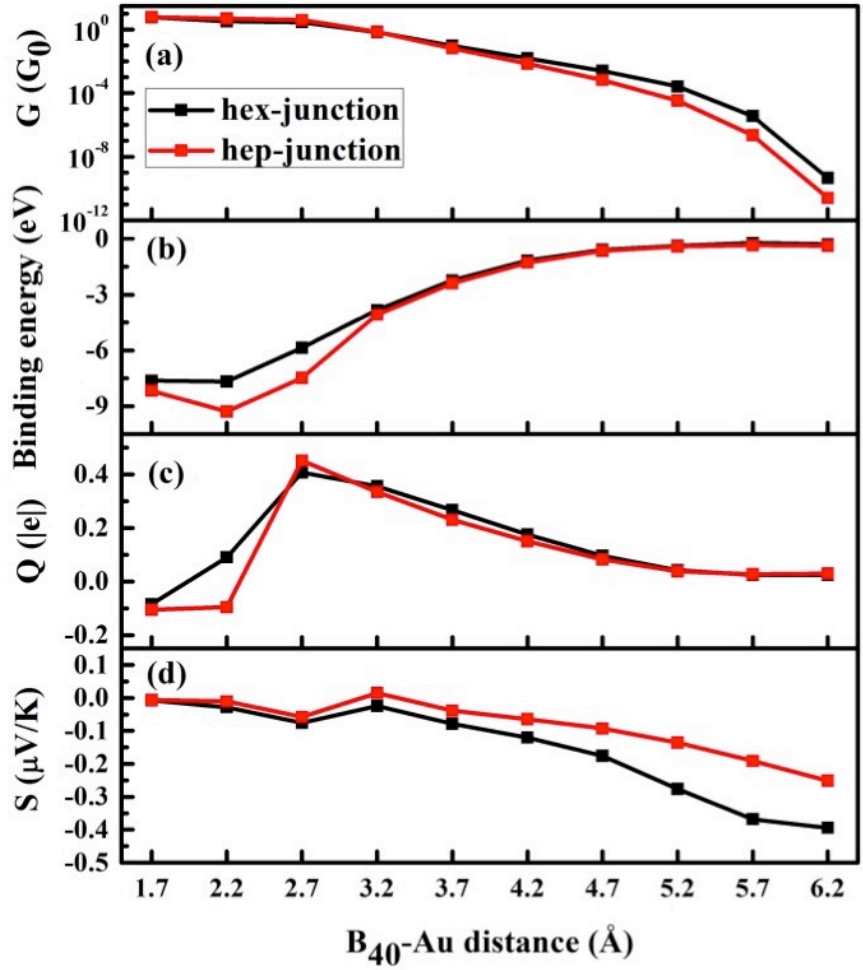

**Figure 3.** (a) Conductance of the two types of Au-B-Au junctions, (b) binding energy of the junctions, (c) Mulliken charge on the $B_{40}$ molecule, and (d) Seebeck coefficient (S) at room temperature, all as functions of the $B_{40}$-Au distance during the stretching process. Black squares represent the hex-junction, while red circles represent the hep-junction.

In the B-Au distance range of 1.7-2.7 Å, where the binding energy is lower than -5.5 eV, the conductance exhibits a "plateau", with values between 2.91-5.98 $G_0$. The plateau indicates that a chemical bond between the electrodes and the $B_{40}$ molecule is formed. This is the contact regime.[10,22] During the stretching process, the contact breaks at the B-Au distance of ~3.7 Å, as suggested by the evolution of the binding energy and the exponential drop in conductance. This is the tunnelling regime.

The distance between two heptagonal faces of a single $B_{40}$ molecule is ~0.37 Å longer than the distance between two hexagonal faces. With the same B-Au distance, the distance between electrodes is smaller for the hex-junction, and it is easier for electrons to tunnel through. Interestingly, in the contact regime, the conductance of the hep-junction is higher than that of the hex-junction (Figure S19). This is simply a matter of competition between two factors as to which one dominates the transport: the charge injection rate or the molecule's ability to scatter electrons. In spite of the longer tunnelling distance, the charge injection has more influence on the transport in the contact regime, leading to higher conductance in the hep-junction, as one more B atom on each side of the $B_{40}$ molecule is in direct contact with the electrodes. In the tunnelling regime, however, the conductance of the hep-junction is lower. This is because the rate at which the electrons tunnel through from the electrodes decays less between two hexagonal faces (smaller diameter) on the $B_{40}$ molecule, resulting in higher conductance for the hex-junction in the tunnelling regime. The Au-B distance of 3.7 Å is not only a point which defines the transition from contact to tunneling, but also a crossover point from the regime in which the transport is dominated by charge injection at contact to the regime in which the transport is dominated by the scattering at the molecule.

The binding energies of the junctions corresponding to the B-Au distances are shown in Figure 3(b). The binding energy reaches its minimum at the B-Au distance of 2.2 Å for both types of junction. (A higher resolution of the binding energy between the B-Au distances of 1.7 Å and 2.7 Å is shown in Figure S19.) The energetically preferred B-Au distance for the hex-junction is 2.0 Å, and for the hep-junction, it is 2.2 Å. At such a B-Au distance, the conductance is 4.59 $G_0$ and 4.86 $G_0$ for the hex-junction and the hep-junction, respectively. As the $B_{40}$-Au distance is shorter than 3.2 Å, the binding energies for both types of junction approach around -5.5 eV and below, indicating that the couplings between the $B_{40}$ molecule and the electrodes have become stronger. The transmission without pronounced peaks in the vicinity of $E_F$ is the direct result of such strong coupling. Mulliken charges on the $B_{40}$ molecule are shown in Figure 3(c). During the stretching process, the Mulliken charges and binding energy approach zero when the B-Au distance becomes larger. The Mulliken analysis shows that both types of junction are positively charged unless the B-Au distance is very close.

The Seebeck coefficient is shown in Figure 3(d). Except for the hep-junction at the B-Au distance of 3.2 Å, the $S$ is negative at all stages of the elongation process, indicating that the low-bias transport is dominated by the LUMO of the molecule at these B-Au distances. It is worth mentioning that the $S$ of $B_{40}$ is 2–3 orders of magnitude smaller than that of $C_{60}$ fullerene,[22,23] and not as sensitive to the contact distance.

**B. Tuning the transport properties by doping.** The conductance of a $C_{60}$-based junction can be tuned by trapping a single atom inside the C cage. We also studied the transport properties of molecular junctions based-on endohedral borospherenes: M@$B_{40}$ (M = Ca, Sr, Y, $H_2O$) with B-Au distance of 3.2 Å. The structures of single endohedral molecules were optimized first. The total energy of the $H_2O$@$B_{40}$ molecule is 0.596 eV lower than the sum of energies of the free $H_2O$ molecule and the $B_{40}$ molecule, indicating that the $H_2O$@$B_{40}$ molecule is a stable molecule.

Then, the optimized molecules are bridged between two Au electrodes. The double-zeta plus polarization basis set for dopant atoms was used. The rest of the parameters in the calculations remained the same.

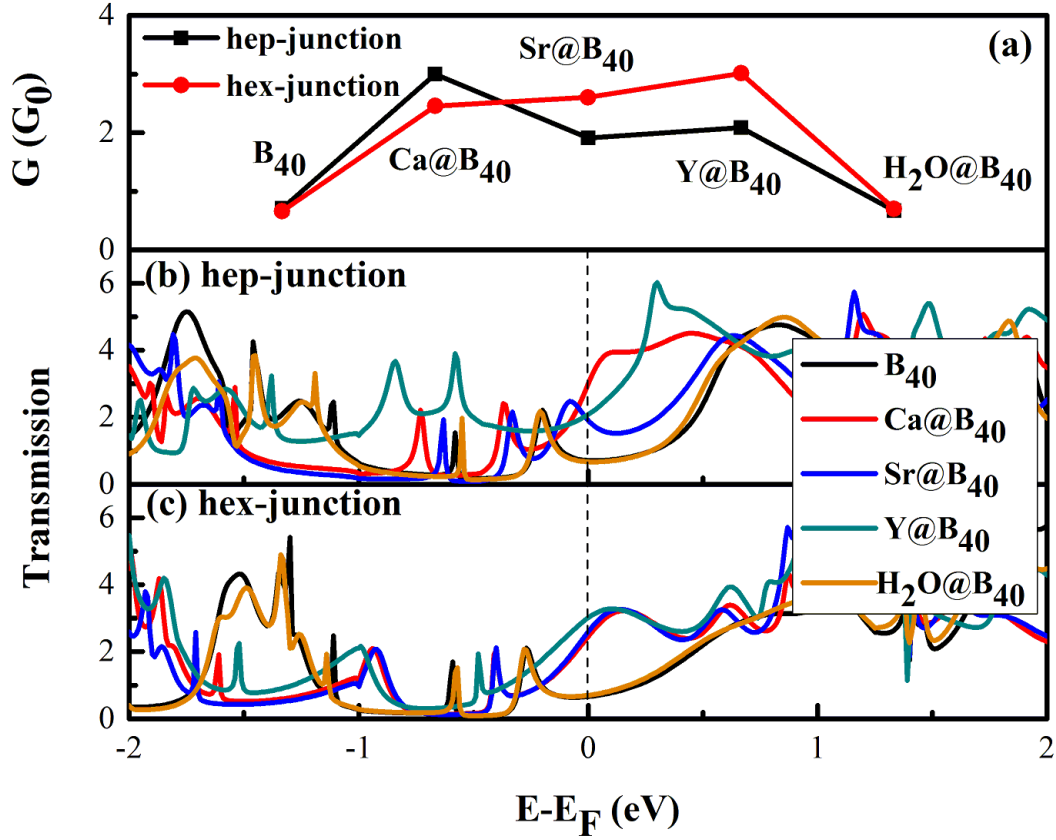

**Figure 4.** (a) Conductance for $B_{40}$-, $Ca@B_{40}$-, $Sr@B_{40}$-, $Y@B_{40}$-, and $H_2O@B_{40}$-based junctions with two coupling geometries: the hex-junction (black squares) and the hep-junction (red circles); (b) and (c) transmission as a function of energy at zero bias for the doped junctions with the two types of coupling geometry. Black, red, blue, cyan and orange represent transmission of $B_{40}$-, $Ca@B_{40}$-, $Sr@B_{40}$, $Y@B_{40}$-, and $H_2O@B_{40}$-based junctions, respectively.

The conductance of a $B_{40}$-based junction is dependent upon the species inside the molecule and the coupling geometry. It increases after metallic doping and changes slightly after $H_2O$ molecule doping, as shown in Figure 4(a). From Figure 4(b) and (c), the transmission of an $H_2O@B_{40}$ junction is similar to that of a $B_{40}$ junction for the two types of contact geometry. For the doped hep-junction, however, it is obvious that the transmission peaks shift downwards after metallic doping (Ca, Sr, and Y). Also, the doping results in a broadening of the LUMO. The $Y@B_{40}$ junction has the highest LUMO peak, while the one for the $Ca@B_{40}$ junction is broadened and spreads over the vicinity of $E_F$, resulting in the highest conductance among the doped junctions. Interestingly, the HOMO splits into two peaks after Sr doping. As can be seen, the broadened HOMO for the $Sr@B_{40}$ junction partially lies on the $E_F$, making it the only metal-doped junction with hole transport at low bias. For the metal-doped hex-junction, the transmission peaks shift downwards without HOMO splitting. All the LUMOs are broadened, with the one for $Y@B_{40}$ being the highest at $E_F$, leading to the conductance of $Y@B_{40}$ being the highest for a doped hex-junction.

To understand the evolution of the conductance with doping, we proceeded by calculating the HOMOs and LUMOs of the five single molecules. The HOMO-LUMO gap for a single $B_{40}$ molecule is 1.76 eV, in agreement with the 1.77 eV in ref 48. The HOMOs and LUMOs are more delocalized after metallic doping and change slightly after $H_2O$ doping (Figure S20), suggesting that the metal-doped molecules tend to be more conductive in molecular junctions. With the electrodes present in the junction, it is useful to visualize the scattering states at $E_F$ that are transmitted through the junction. The scattering states are eigenstates (eigenchannels) of the transmission matrix in Eq. (2). These states characterize the electron transport through the transmission eigenchannels.[31] The charge transfer effects are included, and this information would not be available in the calculations of the free molecule without the Au electrodes. At zero bias, the sum of eigenvalues for each eigenchannel at $E_F$ is the conductance of the junction. For the doped junctions that we studied, 3 eigenchannels dominate the transport properties, and the corresponding scattering states are plotted in Figure 5. Only the scattering states projected onto the bridged molecule are plotted. It can be seen that the molecular scattering states are separately distributed on part of the B atoms. After $H_2O$ doping, the changes in molecular scattering states are negligible. This is why the conductance of the $H_2O@B_{40}$ junction is not changed too much after doping. After Ca, Sr, and Y doping, however, the scattering states change significantly. They are more delocalized, at least in one eigenchannel that was plotted, being almost distributed over the whole bridged molecule in that eigenchannel(s), leading to higher conductance.

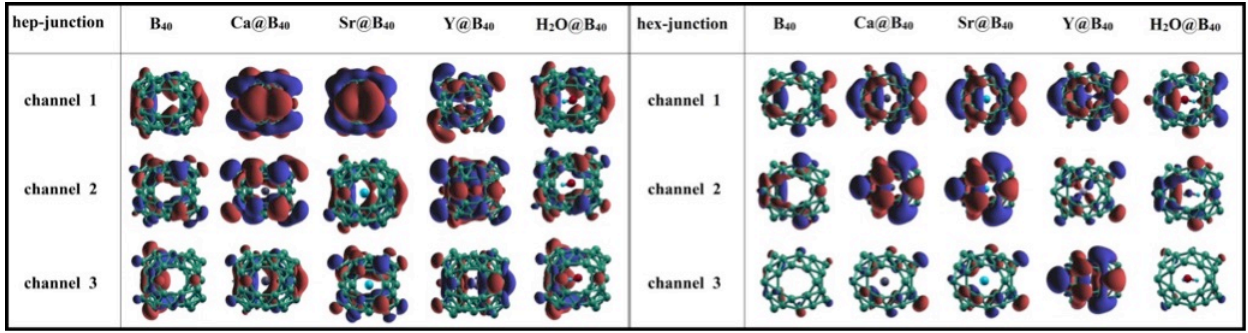

**Figure 5.** Visualization of the three most transmissive eigenchannels (incoming from the left electrode) at the EF for the $B_{40}$, $Ca@B_{40}$, $Sr@B_{40}$, $Y@B_{40}$, and $H_2O@B_{40}$-based junctions. Electrodes are not plotted. The isosurfaces of the eigenchannels are colored according to the phase and sign, with the positive/negative real part being colored in red/blue. The scattering states are plotted with the same isovalue to make them comparable.

Comparing the scattering states in the three different eigenchannels in the $Ca@B_{40}$ hep-junction and hex-junction, those in the second and third eigenchannels are comparable, while the scattering states on the first eigenchannel are more delocalized in the hep-junction. Therefore, the hep-junction has higher conductance. This is not the case in the Sr@$B_{40}$ and Y@$B_{40}$ junctions. From Figure 5, in the Sr-doped junctions, there is only one eigenchannel with the scattering states distributed over the whole molecule in the hep-junction, while there are two eigenchannels with the scattering states distributed over the whole molecule in the hex-junction. In the Y-doped junctions, although there are two eigenchannels with the scattering states distributed over the whole molecule, they are less delocalized compared with the first and third eigenchannels in the hep-junction. As a result, after Sr or Y doping, the hex-junction is more conductive than the hep-junction at the B-Au distance of 3.2 Å.

IV. CONCLUSIONS

In conclusion, we theoretically demonstrate that the $B_{40}$-based junction is highly conductive compared to the $C_{60}$-based junction. For the energetically preferred geometries, the conductance of Au-$B_{40}$-Au junctions can be as high as several times that of comparable Au-$C_{60}$-Au junctions. This is another material where a low-dimensional allotrope reveals distinctive electronic properties from those of the pure bulk. Pure Boron in bulk form is an electrical insulator at room temperature. Unlike single-molecule junctions based on π-conjugated molecules or $C_{60}$ fullerene, the number of conduction channels in a $B_{40}$-molecule junction is less than the number of B atoms in direct contact with the electrode, due to the unique electronic structure of $B_{40}$. Moreover, we have found that the thermopower of $B_{40}$ with gold electrodes is dramatically smaller than that of the Au-$C_{60}$-Au junction. The transport properties of Au-$B_{40}$-Au junctions can be tuned by doping. With a Ca, Sr, or Y atom encapsulated in the $B_{40}$ cage, the conductance at zero bias increases significantly. Our study indicates that the $B_{40}$ fullerene is a new platform for highly conductive single-molecule junctions for future molecular circuits.

**Supporting Information**. Additional information and figures. This material is available free of charge via the Internet at http://pubs.acs.org.

AUTHOR INFORMATION

**Corresponding Author**

*E-mail: xiaolin@uow.edu.au

ACKNOWLEDGMENT

X. L. Wang acknowledges support from the Australian Research Council (ARC) through an ARC Discovery Project (DP130102956) and an ARC Professorial Future Fellowship Project (FT130100778). Computational resources used in this work were provided by Intersect Australia Ltd.

# Theoretical study of transport properties of $B_{40}$ and its endohedral borospherenes in single-molecule junctions

*Chengbo Zhu and Xiaolin Wang\**

*Spintronic and Electronic Materials Group, Institute for Superconducting and Electronic Materials, Australian Institute for Innovative Materials, University of Wollongong, North Wollongong, New South Wales 2500, Australia*

Contents:
Supplementary Fig. S1-S18. The total transmission and individual transmission coefficients as a function of energy for the two geometries with different B-Au distances. The solid black line corresponds to the total transmission, while the others correspond to the contributions of the individual transmission coefficients.
Supplementary Fig. S19. (a) Conductance of the $B_{40}$ junctions (b) binding energy of the junctions, (c) charge on the $B_{40}$ molecule, (d) thermopower at room temperature during the stretching process where the B-Au distance is between 1.7 and 2.7 Å (in steps of 0.1 Å). Black squares represent hex-junctions, while red circles represent hep-junctions.
Supplementary Fig. S20. HOMOs and LUMOs of $B_{40}$, Ca@$B_{40}$, Sr@$B_{40}$, Y@$B_{40}$, and $H_2O$@$B_{40}$ molecules.
Supplementary Fig. S21. The calculated I-V curves of $B_{40}$, Ca@$B_{40}$, Sr@$B_{40}$, Y@$B_{40}$, and $H_2O$@$B_{40}$ junctions with two types of contact geometry.

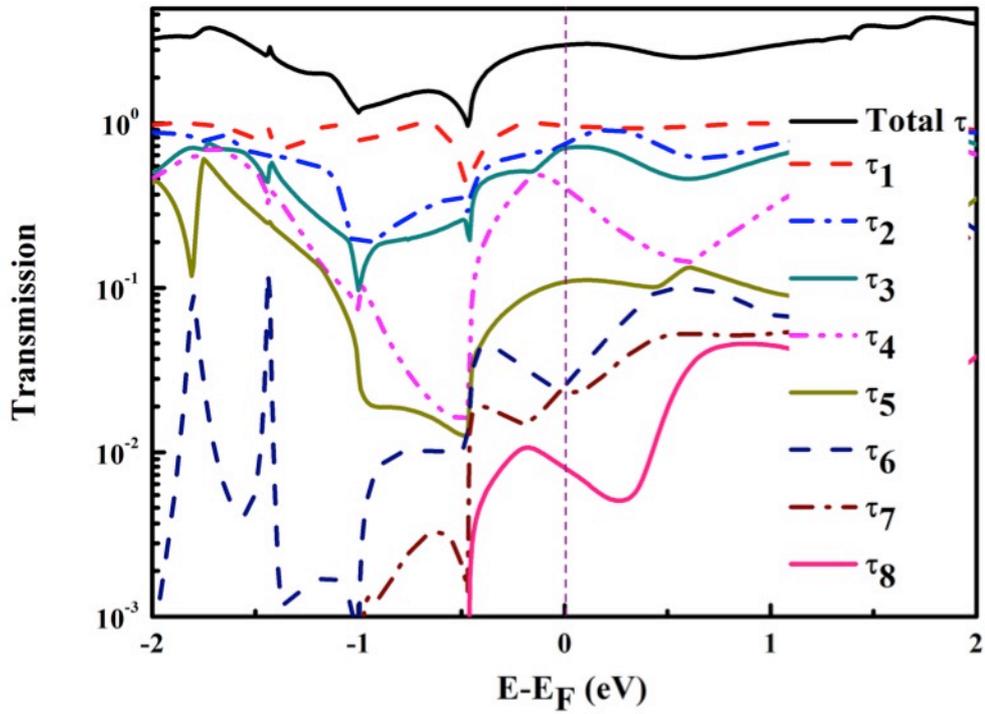

Supplementary Fig. S1. The total transmission and individual transmission coefficients of hep-junction with B-Au distance 1.7 Å.

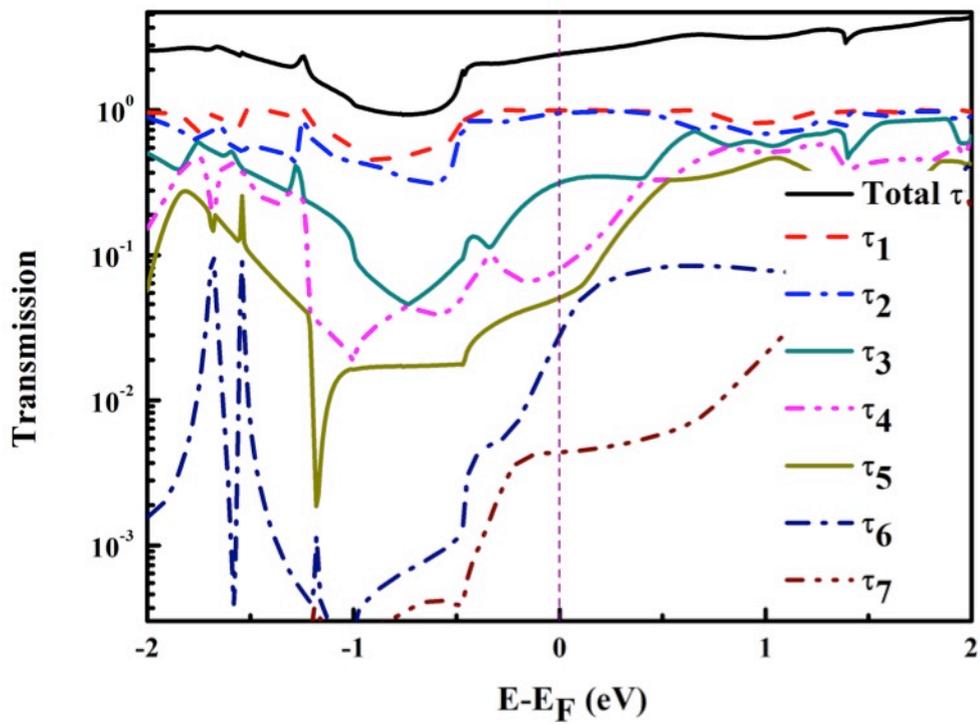

Supplementary Fig. S2. The total transmission and individual transmission coefficients of hep-junction with B-Au distance 2.2 Å.

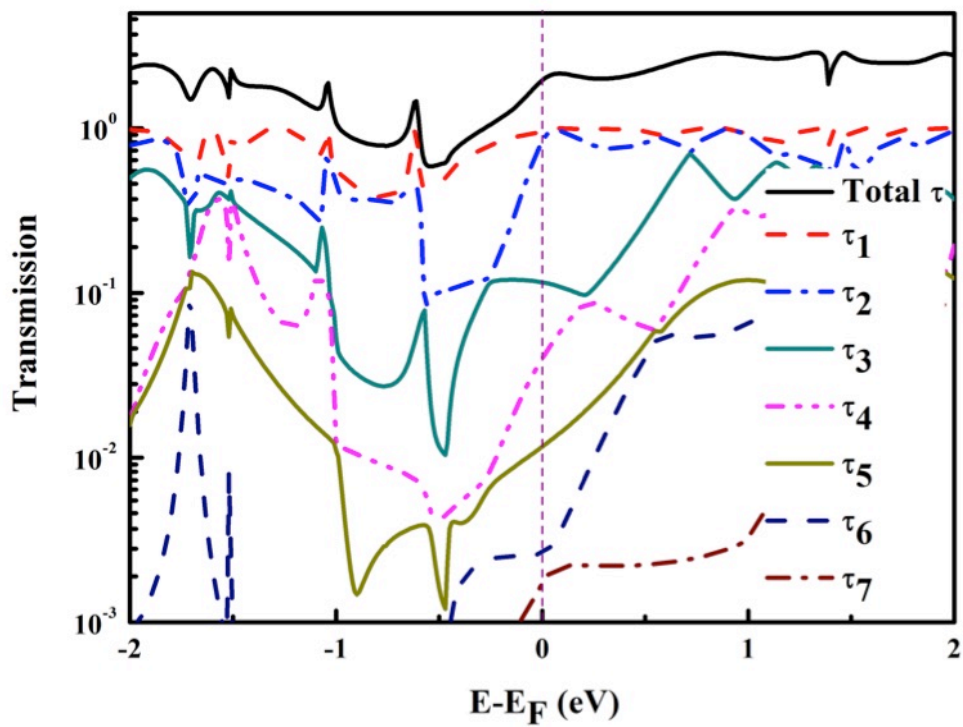

Supplementary Fig. S3. The total transmission and individual transmission coefficients of hep-junction with B-Au distance 2.7 Å.

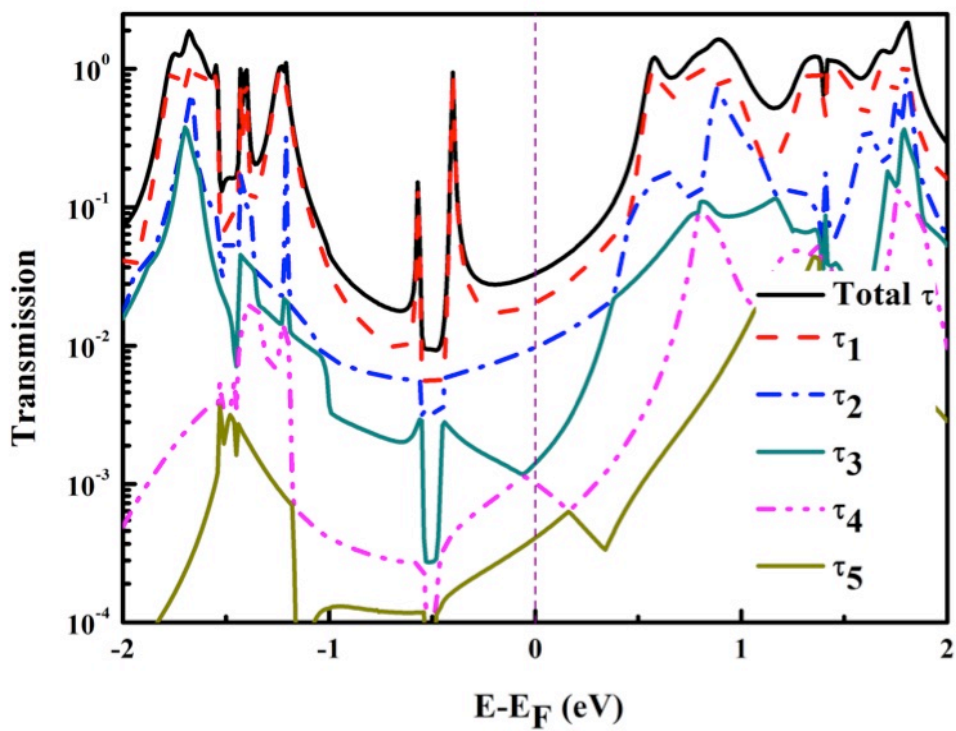

Supplementary Fig. S4. The total transmission and individual transmission coefficients of hep-junction with B-Au distance 3.7 Å.

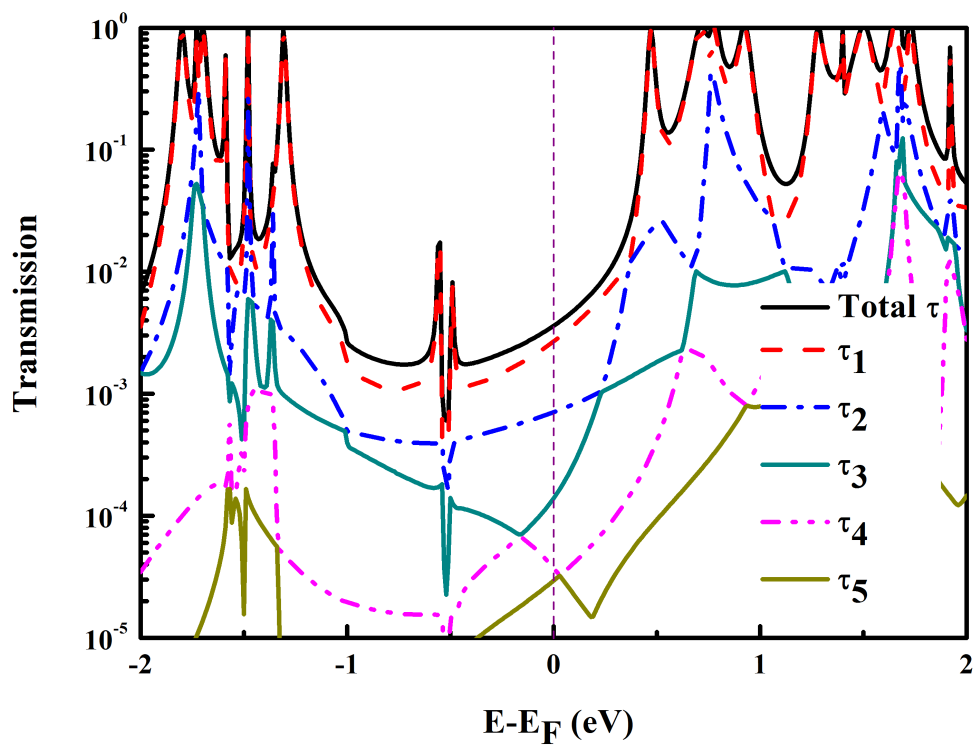

Supplementary Fig. S5. The total transmission and individual transmission coefficients of hep-junction with B-Au distance 4.2 Å

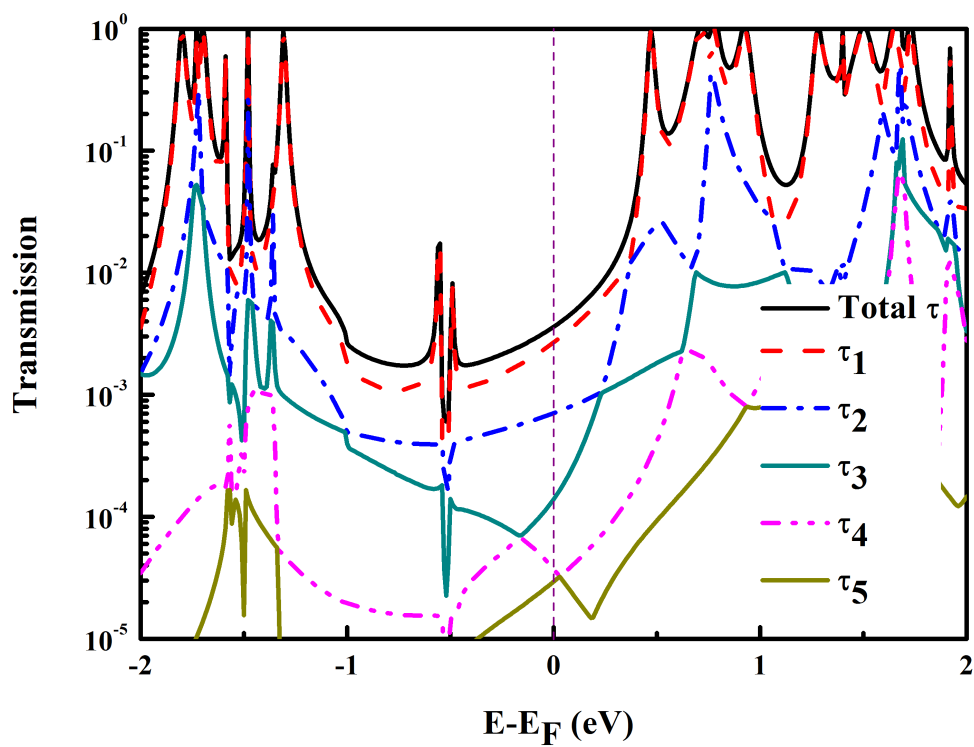

Supplementary Fig. S6. The total transmission and individual transmission coefficients of hep-junction with B-Au distance 4.7 Å.

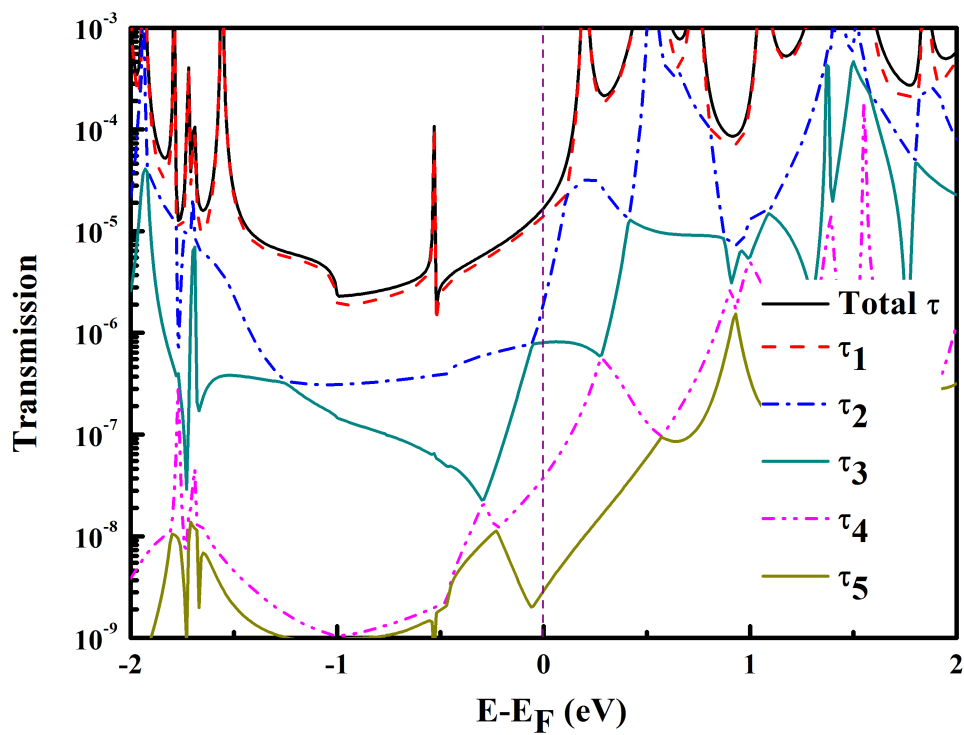

Supplementary Fig. S7. The total transmission and individual transmission coefficients of hep-junction with B-Au distance 5.2 Å.

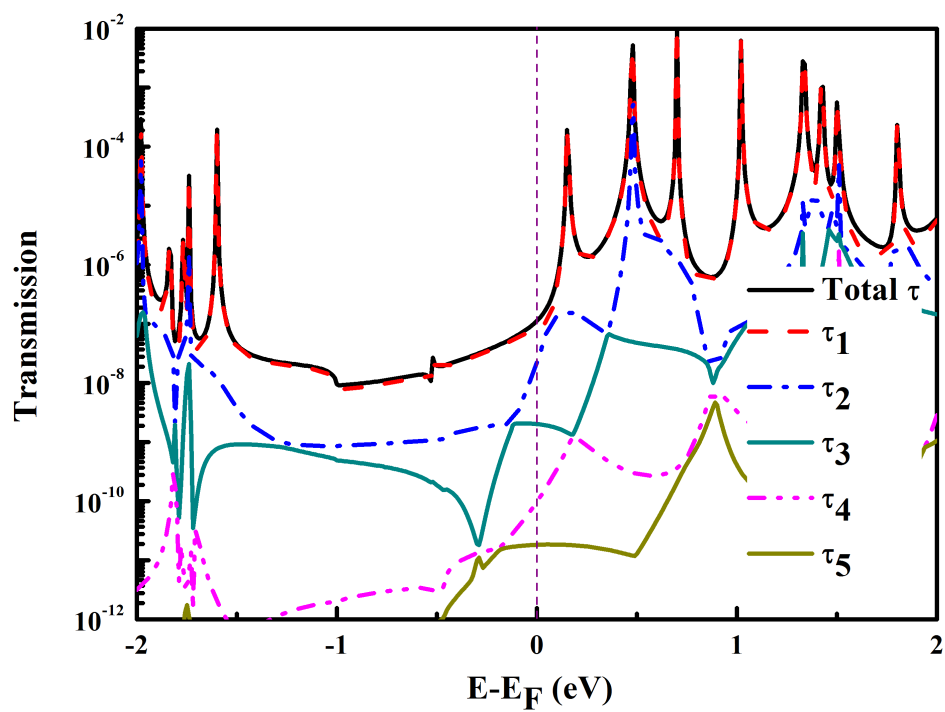

Supplementary Fig. S8. The total transmission and individual transmission coefficients of hep-junction with B-Au distance 5.7 Å.

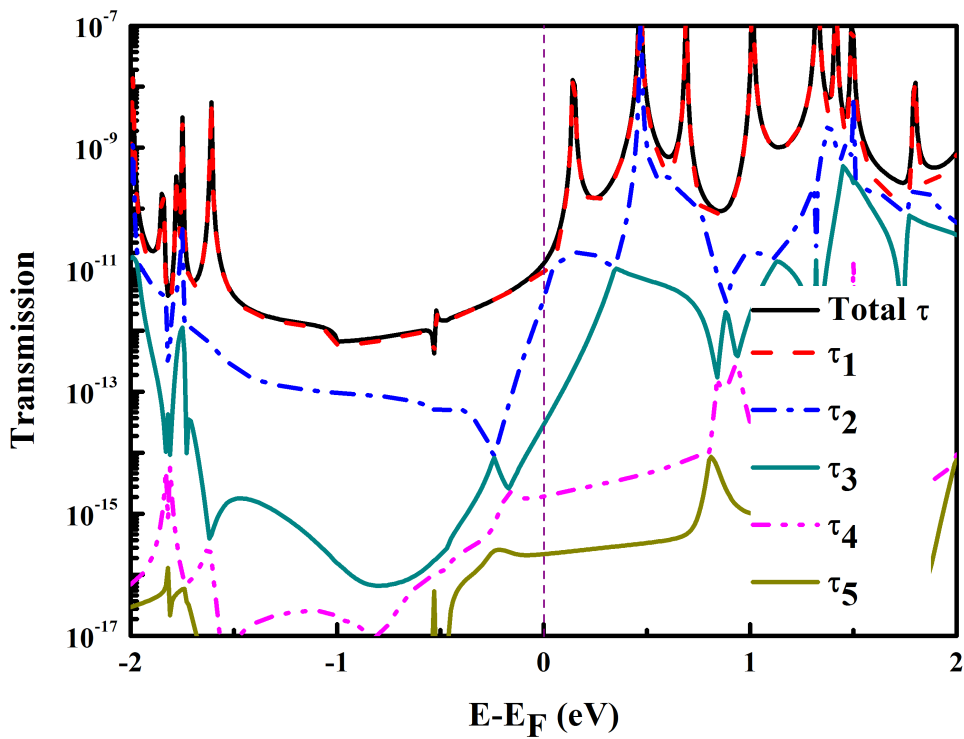

Supplementary Fig. S9. The total transmission and individual transmission coefficients of hep-junction with B-Au distance 6.2 Å.

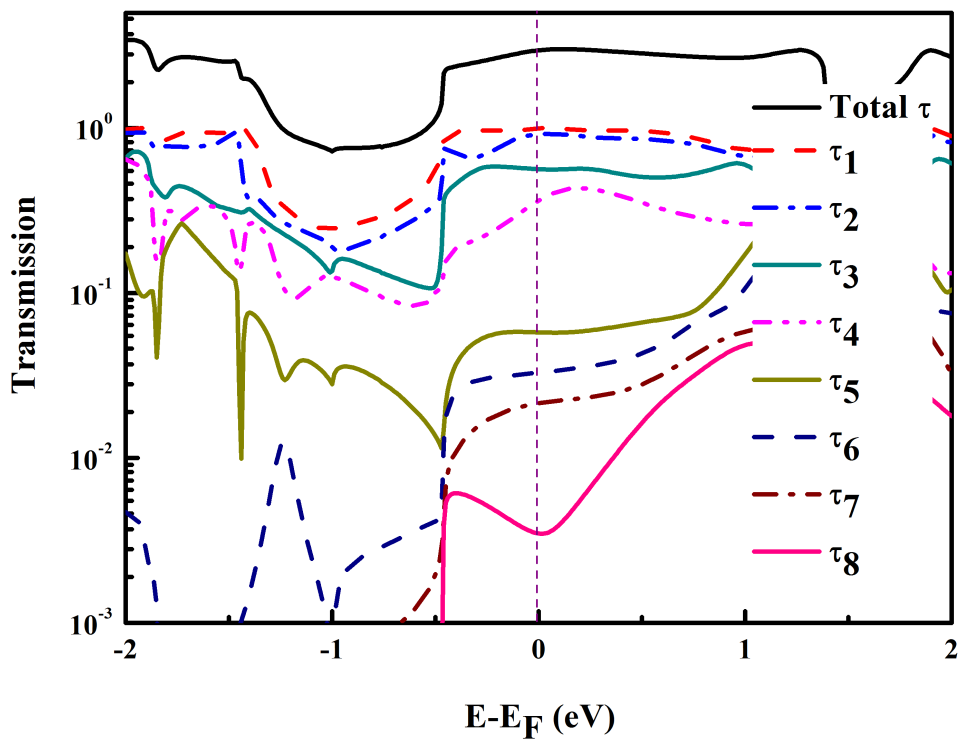

Supplementary Fig. S10. The total transmission and individual transmission coefficients of hex-junction with B-Au distance 1.7 Å.

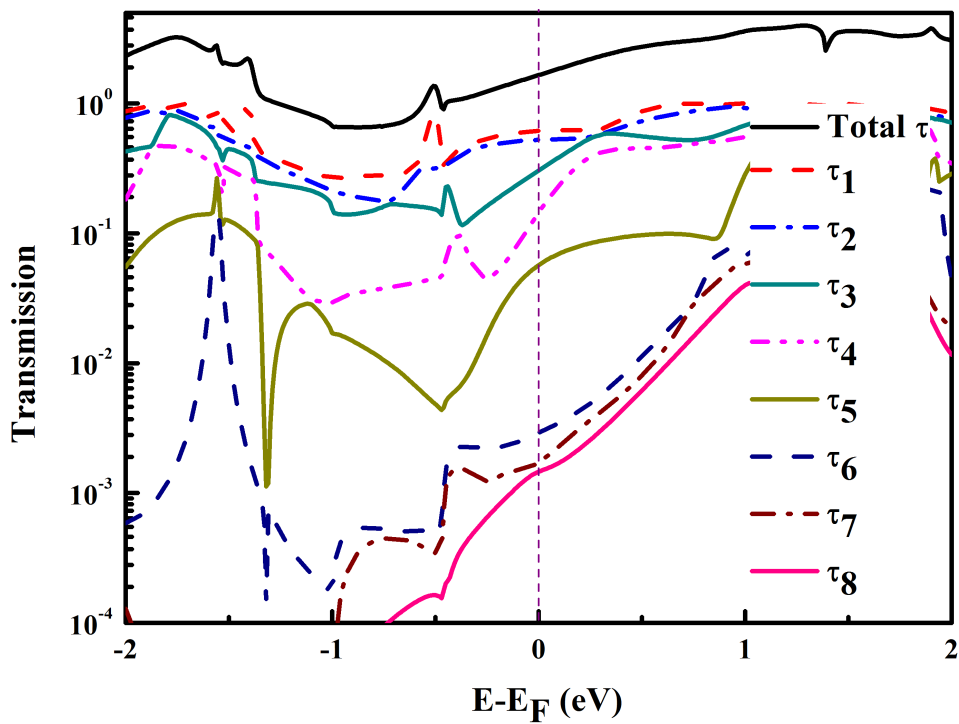

Supplementary Fig. S11. The total transmission and individual transmission coefficients of hex-junction with B-Au distance 2.2 Å.

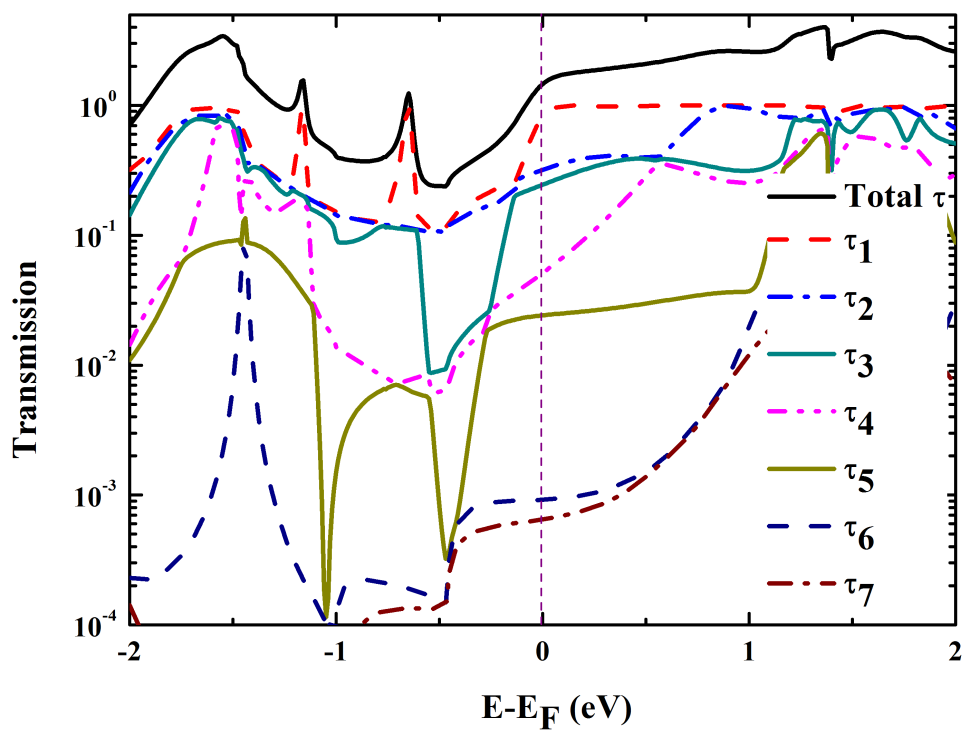

Supplementary Fig. S12. The total transmission and individual transmission coefficients of hex-junction with B-Au distance 2.7 Å.

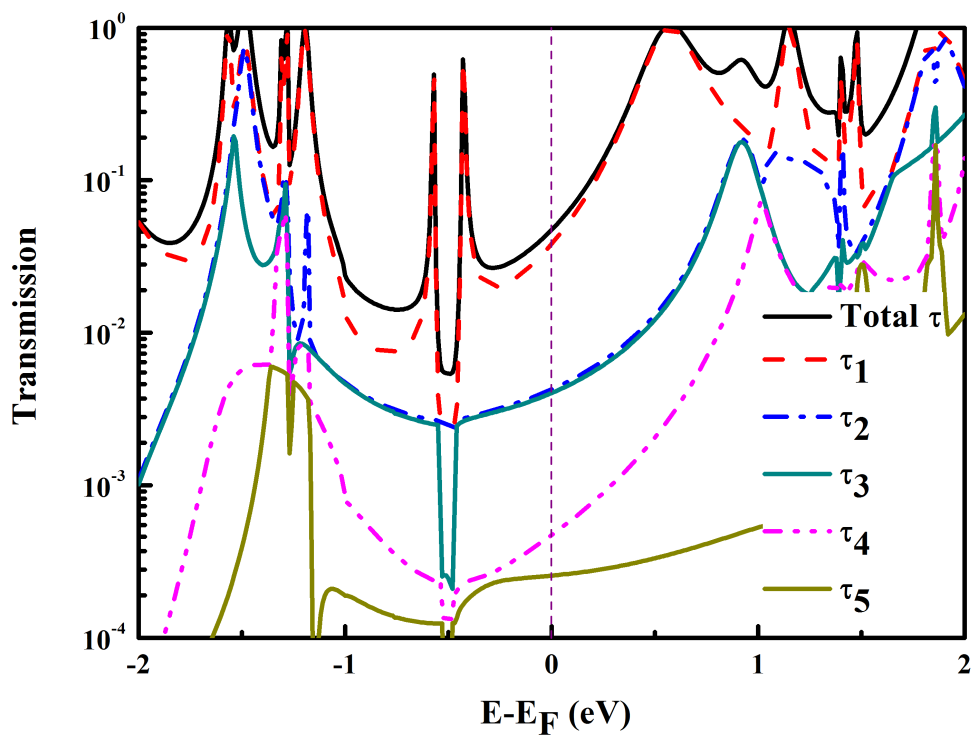

Supplementary Fig. S13. The total transmission and individual transmission coefficients of hex-junction with B-Au distance 3.7 Å.

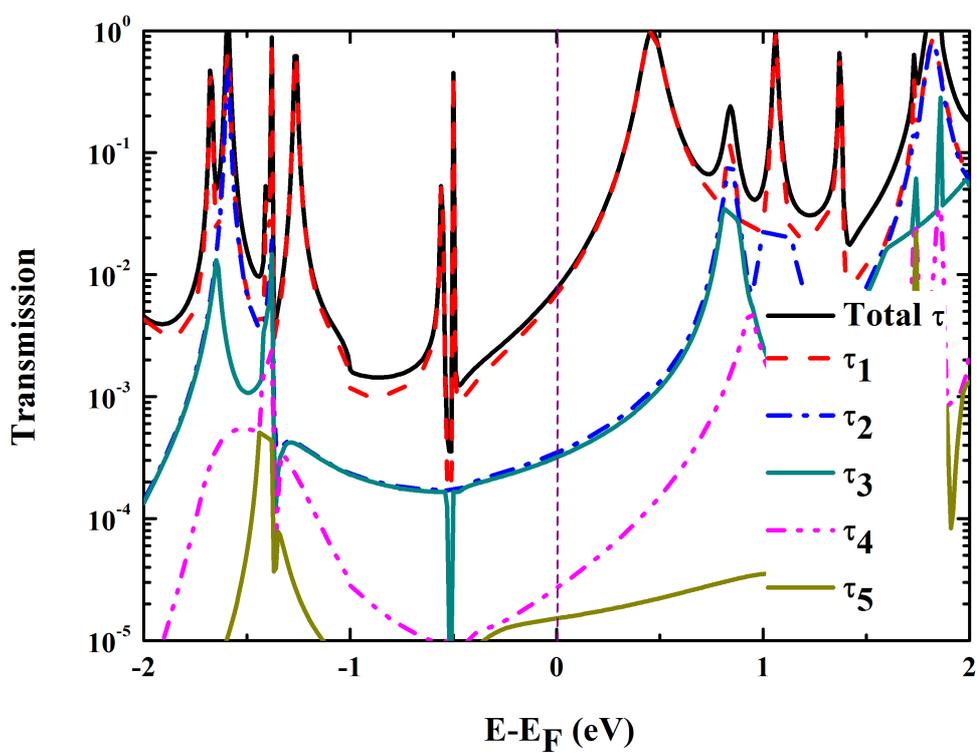

Supplementary Fig. S14. The total transmission and individual transmission coefficients of hex-junction with B-Au distance 4.2 Å.

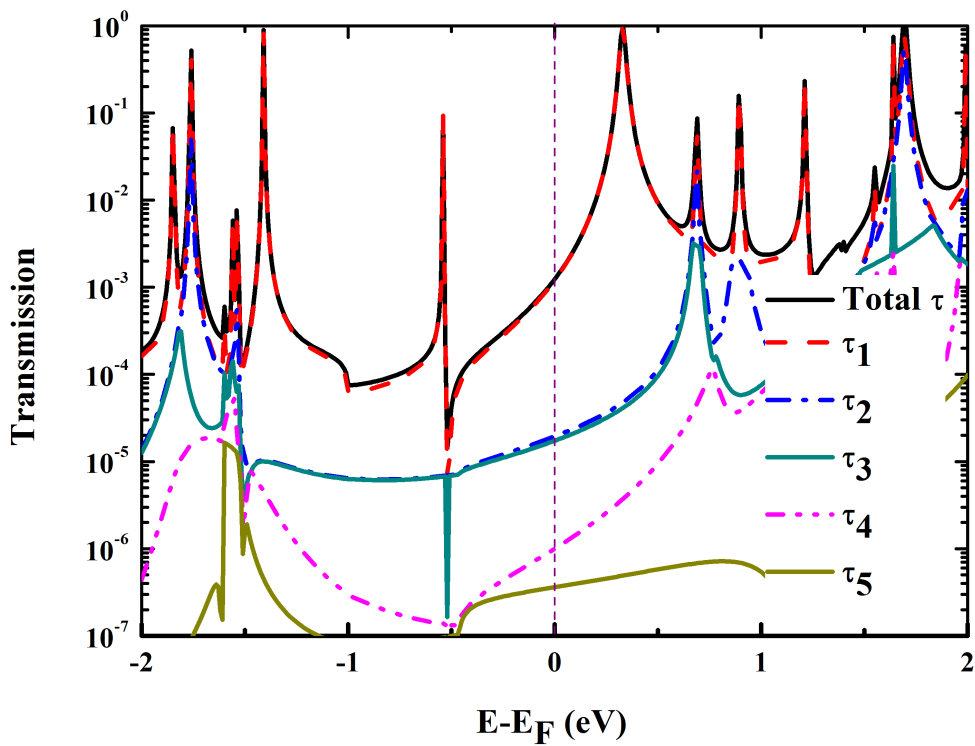

Supplementary Fig. S15. The total transmission and individual transmission coefficients of hex-junction with B-Au distance 4.7 Å.

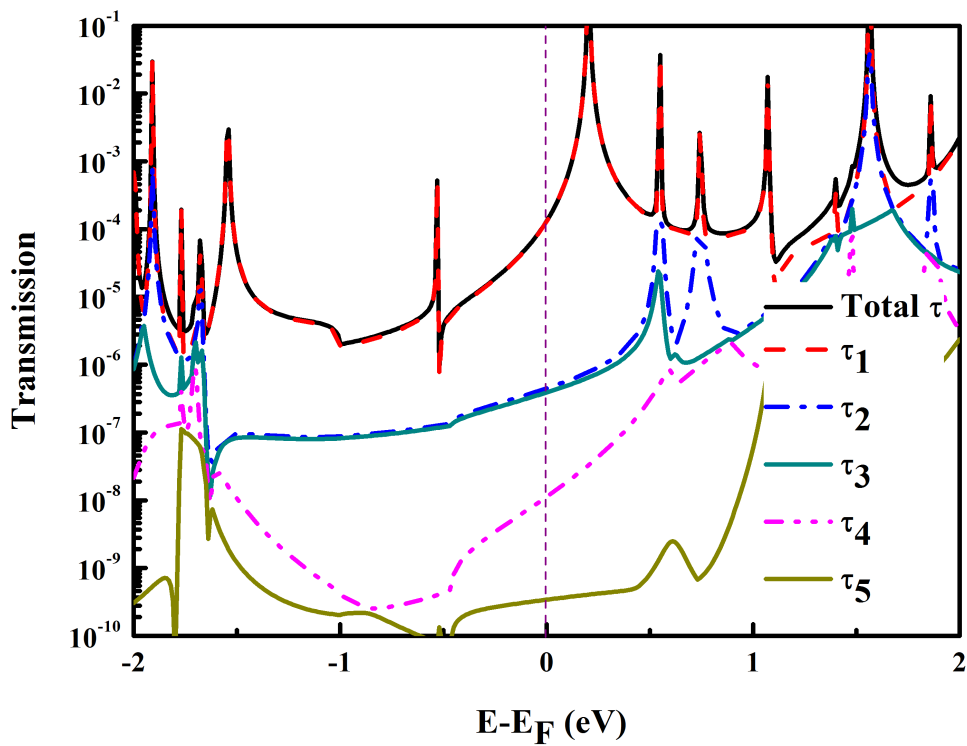

Supplementary Fig. S16. The total transmission and individual transmission coefficients of hex-junction with B-Au distance 5.2 Å.

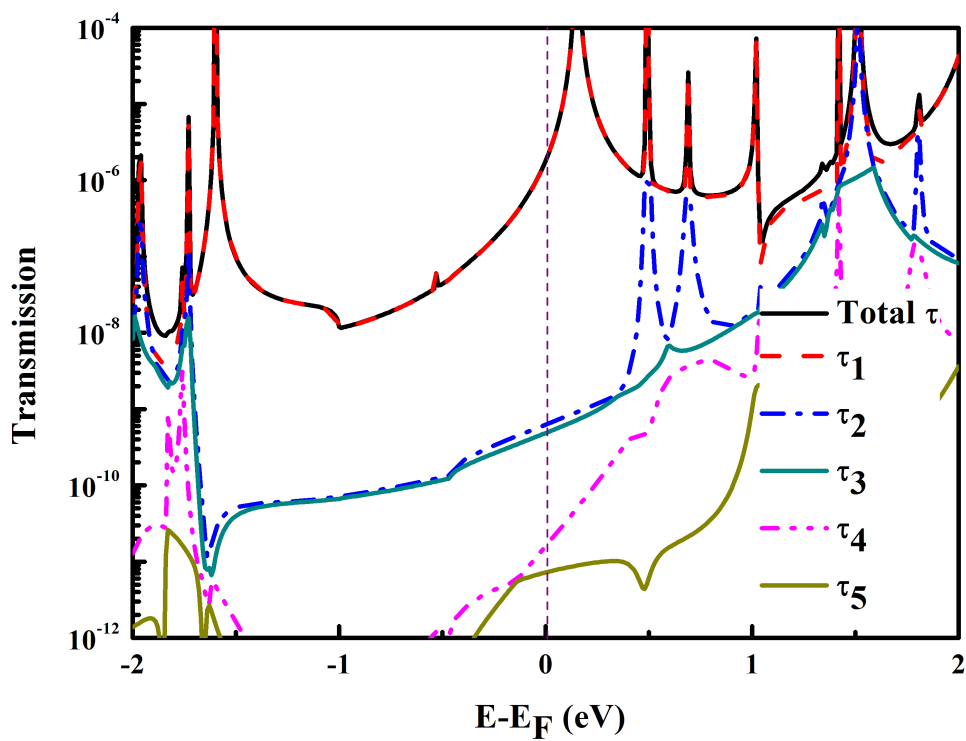

Supplementary Fig. S17. The total transmission and individual transmission coefficients of hex-junction with B-Au distance 5.7 Å.

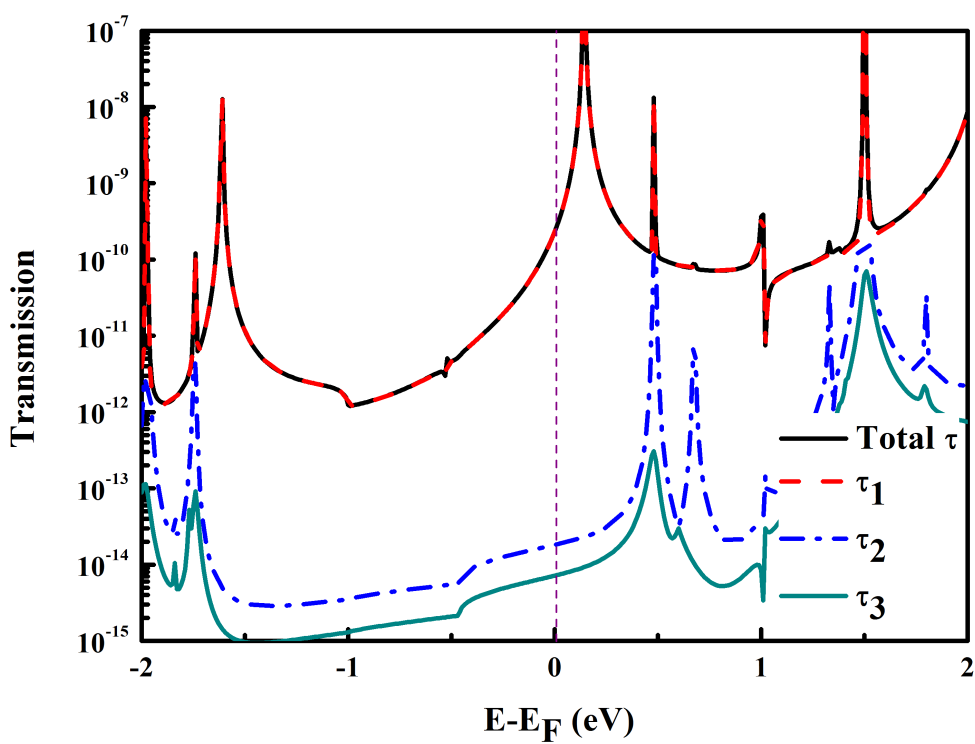

Supplementary Fig. S18. The total transmission and individual transmission coefficients of hex-junction with B-Au distance 6.2 Å.

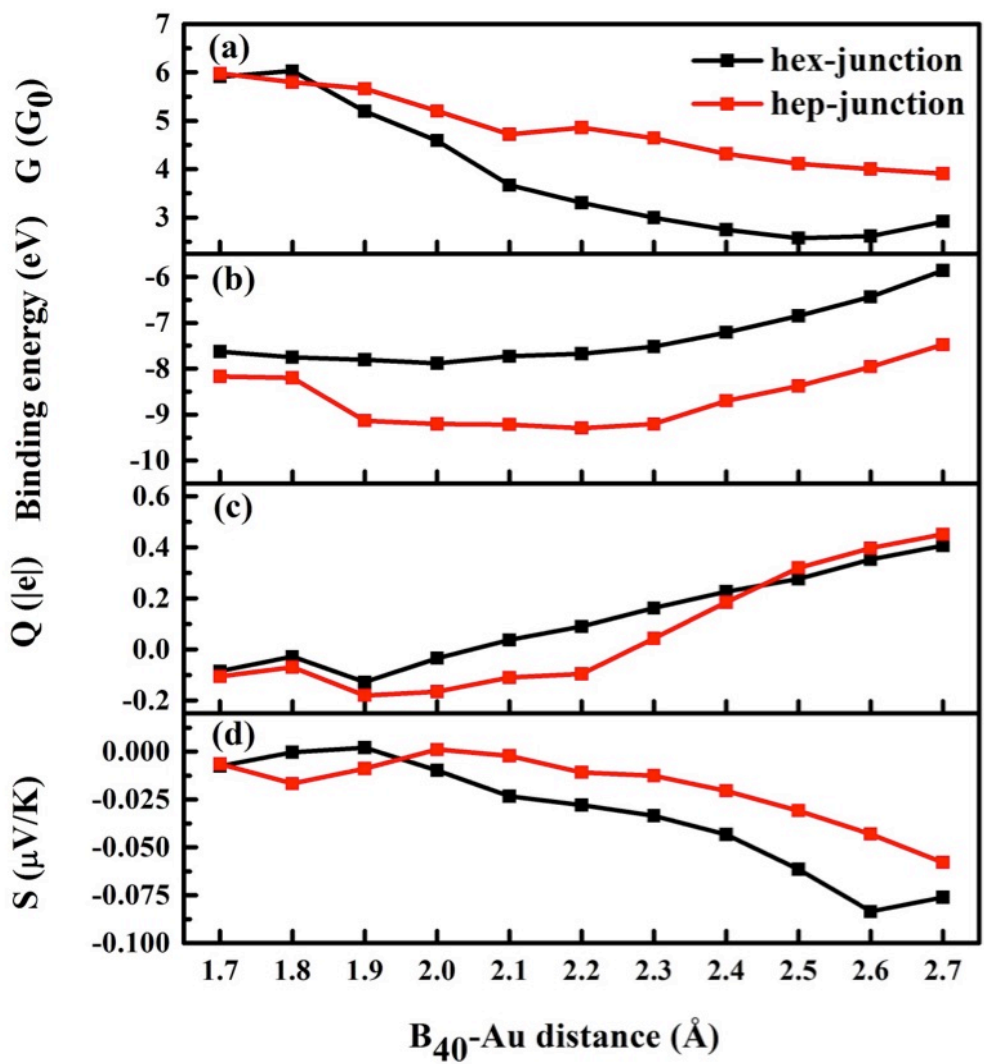

Supplementary Fig. S19. (a) Conductance of the $B_{40}$ junctions (b) binding energy of the junctions, (c) charge on the $B_{40}$ molecule, (d) thermopower at room temperature during the stretching process where the B-Au distance is between 1.7 and 2.7 Å (in steps of 0.1 Å). Black squares represent hex-junctions, while red circles represent hep-junctions.

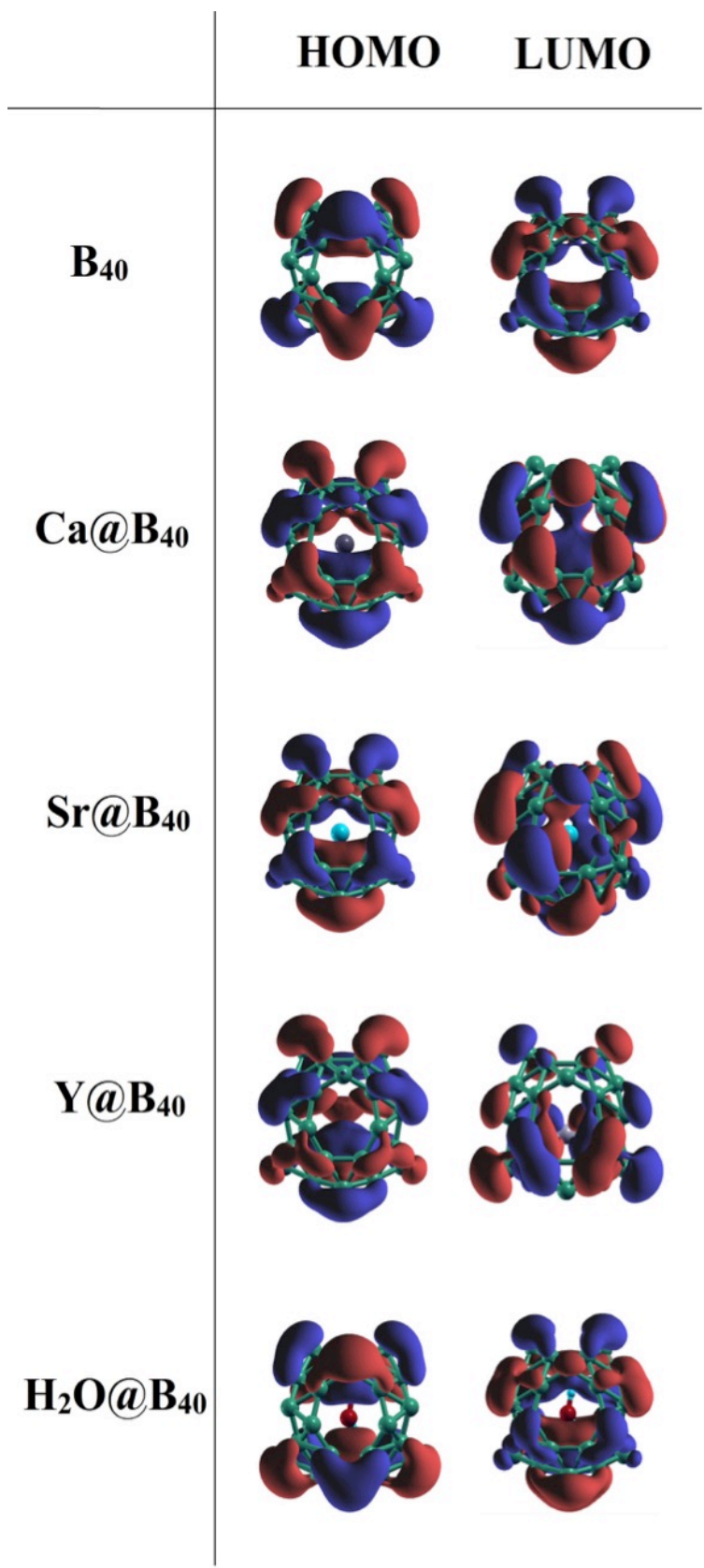

Supplementary Fig. S20. HOMOs and LUMOs of $B_{40}$, $Ca@B_{40}$, $Sr@B_{40}$, $Y@B_{40}$, and $H_2O@B_{40}$ molecules.

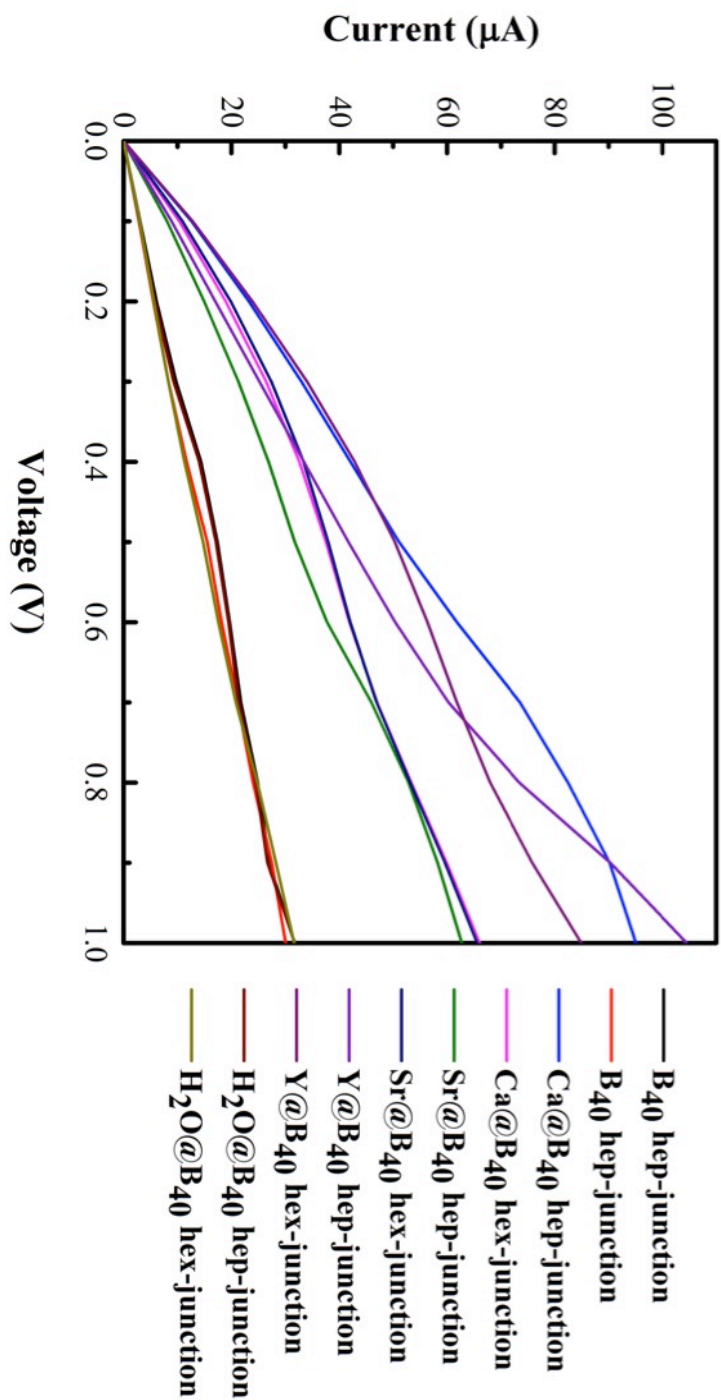

Supplementary Fig. S21. The calculated I -V curves of $B_{40}$, $Ca@B_{40}$, $Sr@B_{40}$, $Y@B_{40}$, and $H_2O@B_{40}$ junctions with two types of contact geometry.